\documentclass[sigconf,9pt]{acmart}

\hypersetup{
           breaklinks=true,   % splits links across lines
           colorlinks=true,   % displays links as colored text instead of blocks
           pdfusetitle=true,  % \title and \author values into pdf metadata
        }

\AtBeginDocument{%
  \providecommand\BibTeX{{%
    \normalfont B\kern-0.5em{\scshape i\kern-0.25em b}\kern-0.8em\TeX}}}

\setcopyright{rightsretained}
\copyrightyear{2022}
\acmYear{2022}
\acmDOI{10.29173/irie480}
\acmConference[IRIE]{The International Review of Information Ethics}{The
International Review of Information Ethics}{Vol. 31, Issue 1, August 2022}
\acmBooktitle{IRIE: The International Review of Information Ethics}
\acmISBN{}

\usepackage{verbatim}
\usepackage[textsize=footnotesize, textwidth=15mm]{todonotes}
\presetkeys{todonotes}{fancyline}{}
\usepackage[english=american,babel=false]{csquotes}
\usepackage{nth}

\usepackage{etoolbox}
\newbool{print}
\booltrue{print}
\usepackage{enumitem}
\newlist{paraenum}{enumerate*}{1}
\setlist[paraenum]{label=\emph{(\arabic*)}}

\begin{document}
%\fancyhead{}

\title{Upgrading the protection of children from manipulative and
addictive strategies in online games: Legal and technical solutions beyond privacy regulation}

\author{Tommaso Crepax}
\email{tommaso.crepax@santannapisa.it}
\orcid{0000-0003-0379-7521}
\affiliation{%
  \institution{LIDER-Lab, Istituto Dirpolis, Scuola Superiore Sant'Anna}
  \city{Pisa}
  \country{Italy}
  \postcode{56127}
}

\author{Jan Tobias M\"uhlberg}
\email{jantobias.muehlberg@cs.kuleuven.be}
\orcid{0000-0001-5035-0576}
\affiliation{%
  \institution{imec-DistriNet, KU Leuven}
  \city{Leuven}
  \country{Belgium}
  \postcode{3001}
}

\begin{abstract}
Despite the increasing awareness from academia, civil society and media
to the issue of child manipulation online, the current EU regulatory
system fails at providing sufficient levels of protection. Given the
universality of the issue, there is a need to combine and further these
scattered efforts into a unitary, multidisciplinary theory of digital
manipulation that identifies causes and effects, systematizes the
technical and legal knowledge on manipulative and addictive tactics, and
to find effective regulatory mechanisms to fill the legislative gaps. In
this paper we discuss manipulative and exploitative strategies in the
context of online games for children, suggest a number of possible
reasons for the failure of the applicable regulatory system, propose an
\enquote{upgrade} for the regulatory approach to address these risks from the
perspective of freedom of thought, and present and discuss technological
approaches that allow for the development of games that verifiably
protect the privacy and freedoms of players.

\end{abstract}


\maketitle
\newpage

\renewcommand{\shortauthors}{Crepax \& M\"uhlberg}
\renewcommand{\shorttitle}{Upgrading the protection of children from manipulative and
addictive strategies in online games}

%{\ } \\
%\vfill
%\vspace{20mm}

% generated with 
% pandoc cleanup.odt -o cleanup.tex

% \date{\today}
% 
% \author{Tommaso Crepax and Jan Tobias M\"uhlberg}
% 
% \title{Upgrading the protection of children from manipulative and
% addictive strategies in online games \\[1ex] \large Legal and technical solutions beyond privacy regulation}
% 
% \maketitle
% 
% \clearpage
% 
% \subsection*{Summary}
% 
% 
% 
% \subsection*{Authors}
% 
% \paragraph{Tommaso Crepax}
% \begin{itemize}
% \setlength{\itemsep}{0pt}
% \item
%   LIDER-Lab, Istituto Dirpolis, Scuola Superiore Sant'Anna, Santa
%   Cecilia 3, 56127 Pisa
% \item
%   \href{mailto:tommaso.crepax@santannapisa.it}{\emph{tommaso.crepax@santannapisa.it}}
% \end{itemize}
% 
% \paragraph{Jan Tobias M\"uhlberg}
% \begin{itemize}
% \setlength{\itemsep}{0pt}
% \item
%   imec-DistriNet, Computer Science, KU Leuven, Celestijnenlaan 200a,
%   3001 Leuven, Belgium
% \item
%   \href{mailto:jantobias.muehlberg@cs.kuleuven.be}{\emph{jantobias.muehlberg@cs.kuleuven.be}},
% \\ \href{https://distrinet.cs.kuleuven.be/people/JanTobiasMuhlberg}{\emph{https://distrinet.cs.kuleuven.be/people/JanTobiasMuhlberg}}
% \end{itemize}
% 
% \clearpage
% \tableofcontents
% 
% \clearpage
\section{Introduction}

\enquote{The capture and re-sale of human attention became the defining
industry of our time}. With these words, Tim Wu~\cite{wu_attention_2017} summarizes
the core mechanics of the \emph{attention economy}. Coined by the
psychologist, economist and Nobel laureate Herbert A. Simon~\cite{simon_bottleneck_1994}, this
term describes an economy where the almost infinite vendors can no
longer sell products at a price, but must make revenue from capturing,
extending, and maximizing user engagement~\cite{commons_immersive_2019}. Attention is
the scarce, precious resource of this extraordinary economy. Product
designers are trained on persuasive behavioral design to \enquote{play
psychological vulnerabilities (consciously and unconsciously)} against
users in the race to grab their attention~\cite{harris_how_2016}, necessitated by
the prominent \enquote{freemium} model, a predatory business model that relies
on the \enquote{harvesting and analysis of user data, in order to predict
and/or manipulate users' preferences, perspectives, and behavior towards
commercial or political outcomes}~\cite{zuboff_surveillance_2019}. As the attention
economy develops into an attention war, it has competing contenders, an
ever changing arsenal of weapons, it brings harm to the civilian
population, and it faces a growing
resistance.

\paragraph{The Contenders.} Facebook, Twitter,
Instagram,\footnote{Recent revelations on
  Facebook's own internal research on effects of Instagram on teenage
  girls: \enquote{Congress grills Facebook exec on Instagram’s harmful
effect on children}, The Guardian, 2021-09-30:
  \href{https://www.theguardian.com/technology/2021/sep/30/facebook-hearing-testimony-instagram-impact}{https://www.theguardian.com/technology/2021/sep/30/facebook-hearing-testimony-instagram-impact}}
Google, but also Epic Games, Blizzard, EA mobile, Roblox, Nintendo,
Nyantic, these gigantic corporations and their advertising networks are
caught in a zero-sum race for users' finite attention, constantly forced
to outperform their competitors using increasingly persuasive
techniques~\cite{harris_how_2016}.

\paragraph{The Weapons.} The contenders use old and new marketing strategies
to harvest attention~\cite{wu_attention_2017}. From Pavlov's experiments to condition
dogs' behavior, to Fogg's persuasive technology~\cite{fogg_persuasive_2002}, behavioral
and neuro-sciences have fed marketing research with deeper knowledge to
understand and condition consumers' behavior.\footnote{Matthews
  clarifies: \enquote{Neuromarketing is simply an expected recruitment of an
  available technology that widens an already impressive suite of
  existing techniques for hidden persuasion}~\cite{matthews_neuromarketing_2015}.}
Developers introduce addiction-by-design~\cite{schull_addiction_2012} and
compulsion-by-design~\cite{kidron_disrupted_2018} in their services, by applying
nudges,\footnote{Cf. \enquote{Nudge: The Final Edition} by
  Thaler and Sunstein, as well as ongoing works of the \emph{5Rights
  Foundation}, \url{algotransparency.org},
  and Tristan Harris.} sludges, dark patterns, and algorithms that
produce personalized content. A mixture of big data analytics, machine
learning and neuromarketing boost old, and create new marketing
strategies. Owning data -- despite the debate of what data ownership
means in legal terms~\cite{duch-brown_economics_2017} --
drives the entire digital ecosystem, as data is the means, the
informational source to develop systems that capture and retain
attention. If attention is the oil, then data is the drill to extract
it.

\paragraph{The Civilians.} Strategies designed to drive and extend user
engagement are central to many of the online services that
\emph{children} use. This is as true for online games as it is for
social media, online streaming services, or search
engines~\cite{kidron_disrupted_2018,council_deceived_2018}. Children are one third of internet users and,
as a substantial portion of freemium games players, they are caught
unguarded in the war for attention.

As Baroness Beeban Kidron puts it, \enquote{the current asymmetry of power
between the developing child and the most powerful companies in the
world is hardly caring, and certainly not in the child's best
interest}~\cite{kidron_disrupted_2018}. In this paper, we identify the
\enquote{developing child} as players during their middle childhood (ages 7
through 12). This age is significant because it marks the time when, on
one hand, children's highly plastic brain is biologically predisposed to
absorb ever more complex information, while their critical thinking, on
the other, is not sufficiently formed~\cite{blumberg_digital_2019}. As \enquote{godlike technologies} take advantage of children's
\enquote{paleolithic emotions}~\cite{wilson_origins_2017}, the legal, social and ethical
systems nowadays in place are not succeeding at ensuring decent levels
of protection~\cite{livingstone_protection_2018} (cf. Section~\ref{sec:3-legal}).

\paragraph{\ldots{}and the Resistance!} Meanwhile, international
organizations, academia, civil society and many other initiatives
worldwide are devoting efforts into showing the detrimental effects of
gaming as well as to understanding the marketing tactics in the digital
domain -- most notably, in 2018 the World Health Organization added
gaming disorders to the list of addictive behaviors. Responses to this
problem are varied, ranging from individuating manipulative tactics and
developing lists and categorizations of nudges, sludges and dark
patterns,\footnote{E.g., 5Rights Foundations~\cite{kidron_disrupted_2018},
Norwegian Consumer council~\cite{council_deceived_2018}, Center for
  Humane Technology, NeuroRights Initiative, AlgoTransparency project.}
to more general approaches to protect childrens' digital
well-being,\footnote{E.g., BetterInternet4Kids,
  \cite{oecd_children_2021}, Gam(e)(a)ble:
  \url{https://www.gameable.info/}.}
through developing technical controls and Privacy Enhancing Technologies
specifically for children,\footnote{E.g., IEEE Standard
  Association report on Children's Data Governance Applied Case Studies
  at
  \url{https://standards.ieee.org/initiatives/artificial-intelligence-systems/childrens-data-governance.html}
and
\cite{crepax2022information}} or asking the industry to adopt
codes of ethics, and governments to take political and legal
actions~\cite{kidron_disrupted_2018, livingstone_protection_2018,
harris_how_2016}.

Although it is admirable that the problem of digital manipulation is
getting attention from multiple stakeholders, its solution is \emph{time
consuming} and \emph{complex}. Time-consuming, because an exhaustive
solution would need a profound, critical social revision of policies
concerning the endangerment of children for commercial purposes through
data exploitation; complex, because alongside societal pressure and
political changes, technological solutions are needed \emph{now} to
momentarily dab the bleeding with immediately actionable responses. That
is, under the hypothesis that once the player's data finds its way into
exploitative advertising networks, the (ab-)use of this data can no
longer be controlled or supervised. Thus, we discuss technological
approaches to verifiably minimize such data
leakage.

\subsection{This Paper \& Contributions}
\label{sec:1-1-paper}

This paper aims to pave the way for the regulatory solution, and provide
a technological option as a potential immediate fix but also to ease the
enforcement of regulatory approaches. To reach these goals, this paper
discusses the following research questions:

\begin{enumerate}
\item
  Is the current European focus on privacy and data protection adequate
  to protect children from manipulative and addictive neuromarketing
  techniques in online games, or is it necessary to \emph{upgrade} the
  focus to the protection of children's right to freedom of thought?
\item
  What technical solutions can be implemented to support a
  privacy-by-design approach to develop game engines that verifiably
  guarantee that player data cannot be extracted by the game server
  operator?
\end{enumerate}

\subsection{Online Gaming: The Paradise of Child Profiling, Manipulation
and Addiction}

The likelihood that children experience harms, as well as their
severity, are directly linked to the amount and sensitivity of data
being accumulated. A substantial portion of children's
\enquote{datafication}~\cite{lupton_datafied_2017}\footnote{Cf. also
\enquote{Age appropriate design: a code of practice for online services}, Information
  Commissioner's Office (U.K., 2020): \url{https://ico.org.uk/for-organisations/age-appropriate-design-executive-summary-and-standards/}} happens through the generation and collection of
massive amounts of online gaming data~\cite{russell_privacy_2018,
newman_press_2014, sax_getting_2021}. In fact, game
developers wallow in \enquote{dataveillance}~\cite{lupton_datafied_2017},\footnote{Dataveillance is the \enquote{monitoring and
  evaluation of children by themselves or others that may include
  recording and assessing details of their appearance, growth,
  development, health, social relationships, moods, behaviour,
  educational achievements and other features}~\cite{lupton_datafied_2017}.} the process of
continuously monitoring, tracking and evaluating activities happening in
the children's real and virtual world. Profiting from In-App Purchases
(IAPs) or third-party advertising, Free-to-Play (F2P) games depend on
the skill of the developer to persuade users into buying items, clicking
on ads or attract more players: this is done by feeding
hyper-personalized content and manipulation through nudging techniques.

Manipulation's success depends on personalization, which is informed
through heterogeneous data collection and processing. Physical world
data are collected through sensors (cameras, microphones,
accelerometers, GPS receivers) while virtual world data are collected
through a users' interactions with the game
software~\cite{sax_getting_2021, newman_press_2014, kroger_surveilling_2021}. The cycle of
processing, called game telemetry,\footnote{Cf. \enquote{The Platform
Evolution of Game Analytics}, Ben Weber, 2018: \url{https://towardsdatascience.com/evolution-of-game-analytics-platforms-4b9efcb4a093}}
starts with the collection of raw game input. Raw game data subsequently
feeds game analytics -- the process for discovering and communicating
patterns in game data. Finally, game data is connected to single
players, becoming \enquote{player metrics}~\cite{newman_press_2014}. Game
analytics and its use to infer player metrics are mature techniques that
are employed in online games for more than a decade
(e.g.,\cite{hullett2012empirical, seif_el-nasr_introduction_2013}). A
recent survey of this field~\cite{su2021comprehensive} concludes that game analytics \enquote{can provide a
service-oriented decision system through data analysis to guide the
whole game industry, especially for the game publishing analytics, which
can help acquire players, maintain players, and maximize game revenue
effectively.}

Following El-Nasr et al.~\cite{seif_el-nasr_introduction_2013} and Su et
al~\cite{su2021comprehensive}, game analytics
and player metrics are most commonly used during game development to
improve the playing experience, to retain players and maximize revenue,
and for marketing purposes. Player metrics analysis are used to
\emph{observe} and \emph{infer} information off the player:
\emph{physical data} such as reflexes, handedness, dexterity are used to
create a physical profile, or, through further evaluation and scoring,
to infer information about the player's health, such as anomalies in sleeping
patterns, state of general health (enough sleep, food, eye rest) or
existence of specific diseases (epilepsy, abnormal dexterity or
reflexes, neurological diseases, dementia, disability, OCDs,
depression).\footnote{As seen in, e.g., the game
  Minecraft with assumptions made on different levels of completion
  [100\%]~\cite{ringland_autsome_2019}, or in choices of colors as potential
  signs of depression.} Adding levels of abstraction, psychological
mechanisms may be used to extract in real time, or predict
\emph{\enquote{mental data}}, such as feelings, emotions, thoughts: behavioral
preferences,\footnote{Player categorization according
  to Bartle's taxonomy: killer, achiever, socializer, explorer, see e.g.
  \url{https://www.essex.ac.uk/people/bartl01006/richard-bartle}}
sexual preferences,\footnote{From gender of character,
  or customization of characters, through clothes, hairstyle, etc.}
propensity for gaming addiction, and more~\cite{madigan_getting_2015,
ienca_mental_2021}. All such highly sensitive information, the essence of
human weaknesses, become vulnerabilities to exploit by the competitors
in the attention economy. Recently, journalists revealed that detailed
profiling using similar approaches to game analytics is also applied in
online gambling. The investigation highlights a company that maintains
\enquote{detailed personal profiles revealing intimate gambling behaviour. The
files contained 186 separate attributes for a single individual, which
painted a detailed and personal portrait of their gambling behaviour,
including their propensity to gamble, favourite games, and
susceptibility to marketing.}\footnote{\enquote{Investigation
  reveals scale of behavioural surveillance by online gambling firms},
  Clean Up Gambling, 2022: \url{https://cleanupgambling.com/news/cracked-labs}} 

\subsection{Profiling in Gaming}

\subsubsection{Telemetry-Driven Behavioral Profiling}

\begin{quote}
\enquote{The old adage of big data having volume, velocity, variety and
volatility holds very true for behavioral telemetry from games.}~\cite{sifa_profiling_2018}
\end{quote}

While players do volunteer some personal data, the vast majority is
acquired by observation, derived through analysis, or inferred
probabilistically off the \emph{unaware
subject}~\cite{sartor_regulating_2021, ienca_mental_2021,
onnela_harnessing_2016}. The
constant growth of training data volume generated by telemetry-driven
behavioral profiling of thousands -- sometimes many
millions\footnote{See e.g., the number of
  players of Fortnite in 2021 being between 6 and 12 million.
  \url{https://www.ggrecon.com/guides/fortnite-player-count-how-many-people-play/}} -- of
players worldwide feeds supervised machine learning algorithms with
accurate, labeled data classes to create nearly-perfect user profiles;
reinforcement learning algorithms, mimicking humans \enquote{rewards
seeking}
actions~\cite{schultz_neural_1997}, improve at predicting
behaviors by exploiting rewards signals embedded by design in the
software code; unsupervised learning algorithms fed by raw data find
unanticipated users' clusters, affiliations and anomalies. Thanks to
data analytics, AI and machine learning techniques, the game gets to
know the users and exploit what triggers
them.

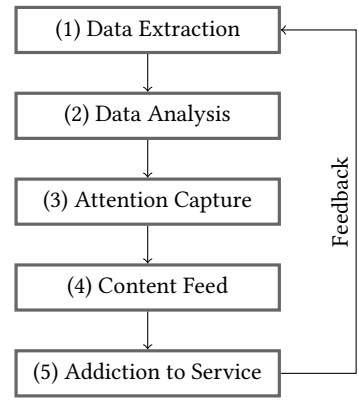
\begin{figure}[ht]
\centering
\begin{tikzpicture}[
squarednode/.style={rectangle, draw=black!60, very thick, minimum size=10mm, minimum width=35mm, minimum height=6mm},node distance=0.5cm
]

\node[squarednode, align=center] (extraction) {(1) Data Extraction};

\node[squarednode, align=center] (analysis) [below=of extraction] {(2) Data Analysis};

\node[squarednode, align=center] (caption) [below=of analysis] {(3) Attention Capture};

\node[squarednode, align=center] (feeding) [below=of caption] {(4) Content Feed};

\node[squarednode, align=center] (addiction) [below=of feeding] {(5) Addiction to Service};

\draw[->] (extraction.south) -- (analysis);
\draw[->] (analysis.south) -- (caption);
\draw[->] (caption.south) -- (feeding);
\draw[->] (feeding.south) -- (addiction);
\draw[->] (addiction.east) -- +(1,0) |- node[pos=0.25,distance=5mm,text depth=-1ex,rotate=90,above]{Feedback} (extraction.east);

\end{tikzpicture}
\caption{Addiction process: from personal data extraction to user addiction.}
\label{fig:addictionprocess}
\end{figure}

The ability to predict which users will turn into long-term players,
social network enablers and/or buyers of in-game content enables an
optimization of Customer Relationship Management, and the tailoring of
game content to the specific profiles of these users~\cite{sifa_profiling_2018}. Much like professionals at chicken
sexing,\footnote{See \enquote{The Lucrative Art of Chicken Sexing},
Pacific Standard, 2018-08-08: \url{https://psmag.com/magazine/the-lucrative-art-of-chicken-sexing}}
shallow profiling algorithms must individuate churners and buyers in as
little time as possible (ca. 20 minutes~\cite{newman_press_2014}). Deep profiling, instead, provides a toolbox for integrating
varied player behaviors and experimenting on potential correlations with
player psychology~\cite{sifa_profiling_2018} through
unsupervised machine learning. The process can be simplified in the
following tasks: \emph{extraction and analysis} of data on user to
understand weaknesses, \emph{capture} of attention through manipulative
tactics, \emph{feeding} ever personalized content until \emph{addiction}
to service is achieved (see Figure~\ref{fig:addictionprocess}).
% \href{https://docs.google.com/document/d/1wvdzJO-wPv7Kr3-p5qb5jdhigDMBak3xxSL9_hnqVuQ/edit\#bookmark=id.20xfydz}

\subsubsection{Psychological Profiling and Neuromarketing}

In such hyper-competitive market for attention, where games' designs are
based on customers' data, telemetry-driven behavioral
profiling\footnote{It is a mix of snapshot, dynamic,
  contextual and spatio-temporal profiling.} is only the
\emph{quantitative} side of the medal. The \emph{qualitative} side is
\emph{psychological profiling}. Marketers exploit psychographics,
user-testing, surveys, focus-groups for customer modeling based on
\emph{self-reported} consumer's emotions, values, personality
traits,\footnote{E.g., D.W. Fisk's OCEAN model, acronym
  for Openness, Consciousness, Extraversion, Agreeableness and
  Neuroticism} lifestyle, opinions, interests, and try to build an
understanding of the psychological state of the player at points in the
future.\footnote{This understanding of the psychological state in the \enquote{human futures market},
  as said in~\cite{zuboff_surveillance_2019}, is the sale of what we will
do next~\cite{alegre_protecting_2021}.}

Psychological profiling based on market research techniques may still
suffer from inaccurate measurements due to qualitative self-reporting
biases in the samples~\cite{wright_history_1997, canli_when_2006, ford_what_2019}. Therefore,
to the problem of precise quantification of mental processes,
neuroscience applied to marketing, although in its
infancy~\cite{ramsoy_building_2019}, provides promising answers~\cite{swan_sensor_2012}: born to study mechanisms
to understand the consumer's behavior~\cite{smidts_kijken_2002}, neuromarketing
exploits physiological-measuring techniques such as eyetracking,
pupilometry, EEG, fMRI, facial coding, sensory marketing, as well as
psychographics~\cite{kenning_consumer_2011}, to measure targeted
people's emotions and feelings~\cite{lindstrom_buy_2010} in reaction to stimuli
(graphic lines, content or ads) at \emph{subconscious} level.

Input from neuromarketing findings are coded by developers into video
games design, translated into more attractive graphic lines, landing
pages and microsites. Stimulated by using an uninterrupted series of
targeted persuasion attempts~\cite{wilson_neuromarketing_2008}, when the
users finally desist from playing, they start receiving timed alerts or
push notifications~\cite{kidron_disrupted_2018} that pull them back into the
game.\footnote{See U.K. Research and Innovation
  (UKRI)'s written evidence for the \enquote{Immersive and Addictive
Technologies}
  report, 2019-08-09: \url{https://publications.parliament.uk/pa/cm201719/cmselect/cmcumeds/1846/1846.pdf}}

\subsubsection{Manipulative and Addictive Tactics}

The manipulative and addictive tactics, designed over neuroscientific
evidence into most digital services available on the Internet, are
irrespective of that there is a high chance (one in three~\cite{livingstone_one_2016-1}) that the user is a child. Although some
manipulative tactics may come from the unintentional, naive use of
well-established persuasive design tactics, other come from the
conscious, malevolent choice of the developers of a service for a child,
or because the service, albeit not for children, is accessed by children
nonetheless -- in fact, most age verification systems fail at determining
age~\cite{pasquale_review_2020}.

There are \enquote{literally thousands}~\cite{harris_how_2016} of \enquote{sticky
features} to
hijack users' minds, divided into \emph{nudge} features, that push users
into behaviors that are in the commercial interest of services, and
\emph{sludge} features, that are barriers to users making decisions in
their own interests~\cite{kidron_disrupted_2018}: tactics range from controlling
menu choices, to exploiting reward loops,\footnote{The
  idea is that of putting \enquote{slot machines in users' pockets,}
paraphrasing Tristan Harris' aticle \enquote{The Slot Machine in Your
Pocket}, Spiegel International, 2016-07-27: \url{https://www.spiegel.de/international/zeitgeist/smartphone-addiction-is-part-of-the-design-a-1104237.html}}
or fears of
missing something important; items can be \enquote{like} buttons to stimulate
social reciprocity, infinite feeds, video autoplay, instant
interruptions, daily login and streaks rewards, no save-no pause
defaults, limited time offers, coins-for-ad watching, burning graphics,
pull to refresh and swipe mechanisms, timed alerts, push notifications,
\enquote{summons} such as buzzes, pings, vibrations, or even the color
red~\cite{sher_framework_2011, kidron_disrupted_2018, sax_getting_2021}.

There is an emerging -- yet underdeveloped~\cite{melzer_towards_2021} -- body of literature specifically on addictive game design that
addresses the interplays among freemium content, ethics and youth. Such
literature shows overall consensus that the freemium model has
negatively impacted on game design~\cite{alha_rise_2020} and~\cite{karlsen_balancing_2021}, for instance through imposing the necessity
of ever increasing predatory~\cite{king_predatory_2018} \enquote{loot
boxes}~\cite{karlsen_game_2020}, mostly due to challenging monetization
mechanics~\cite{karlsen_balancing_2021}. Interviews with informants from game design companies
with different business models showed that freemium companies, as
opposed to premium and indie developers, tend to downplay ethical
responsibilities. This is because, as one freemium developer said, it is
\enquote{literally impossible to launch an ethical game on the App
Store}~\cite{karlsen_balancing_2021}, stressing that freemium developers \emph{need} to
manipulate players into micro purchases. In response to such industry's
need, some solution seekers suggested the use of framework like the App
Dark Design (ADD) to critically evaluate the design of freemium
apps~\cite{fitton_creating_2019}, which, combined with works on identification of technical
controls designed specifically to protect children's privacy~\cite{crepax2022information}, could help game developers in making ethical game
design choices.

\section{The Effects of Manipulative and
Addictive Content on Children}
\label{sec:2-effects}

In this attention economy driven by AI and big personal data analytics,
the impact on users is unprecedented and unforeseeable~\cite{zuboff_surveillance_2019},
but surely significant and severe. In fact, in 2019, the Committee of
Ministers of the Council of Europe produced a declaration \enquote{on the
manipulative capabilities of algorithmic processes} (hereinafter, the
\enquote{Declaration}), where they remark:

\begin{quote}
\enquote{Fine grained, sub-conscious and personalized levels of algorithmic
persuasion may have significant effects on the cognitive autonomy of
individuals and their right to form opinions and take independent
decisions. These effects remain underexplored but cannot be
underestimated. Not only may they weaken the exercise and enjoyment of
individual human rights, but they may lead to the corrosion of the very
foundation of the Council of Europe. Its central pillars of human
rights, democracy and the rule of law are grounded on the fundamental
belief in the equality and dignity of all humans as independent moral
agents.}~\cite{europe_declaration_2019}
\end{quote}

Persuasive design strategies, deployed to maximize the collection of
personal data, disturb children's social, mental and physical
development~\cite{kidron_disrupted_2018}.

Harms include immediate effects on individual consumers as well as
long-term effects on society, and span over a wide range of categories,
from harms to human rights, to physical and psychological harms, through
social harms. The involved rights include positive rights to privacy,
autonomy, and dignity as well as negative rights not to be deceived,
subjected to experiments without consent, or used as a
means~\cite{stanton_neuromarketing_2017}. Personal harms include anxiety,
social aggression, denuded relationships, sleep deprivation, impacts on
education, health and wellbeing~\cite{kidron_disrupted_2018}, psychosocial and
financial harms, withdrawal from real life, heightened attention-deficit
symptoms~\cite{carr_shallows_2020}, impaired emotional and social intelligence,
technology addiction, social isolation, impaired brain development, and
disrupted sleep~\cite{small_brain_2020}. Because of young children's high
brain plasticity, there may be consequences due to the normalization of
gambling~\cite{drummond_video_2018}.\footnote{Henrietta
  Bowden-Jones notes: \enquote{Habits learnt in childhood require significant
  intervention, and that habits formed before the age of nine take
  considerable interventions to change in adulthood. Given that many of
  the techniques used in gambling are deployed both in online games and
  other kinds of services used by children, this should be a source of
  real concern}, from
  \enquote{Video games are pushing children into gambling warns
England’s top mental health nurse}, Health Tech Digital, 2020-02-19:
\url{https://www.healthtechdigital.com/video-games-are-pushing-children-into-gambling-warns-englands-top-mental-health-nurse/}}
Scientific literature is increasingly coming to the conclusion that
addiction is at least a disease of free will~\cite{koob_neurobiology_2016}, and
at worst a brain disease~\cite{leshner_addiction_1997, fenton_free_2017}. As not
only is attention finite, but also the \enquote{bottleneck of human
thought}~\cite{simon_bottleneck_1994}, providing only content that pleases the child's
interests takes their attention away from diverging content and is a
hindrance to develop robust critical thinking. Finally, Small~\cite{small_brain_2020} has demonstrated that \enquote{internet addiction shares features
with substance-use disorders or pathological gambling}.

As an answer to these dramatic risks, there have been some authoritative
recognition of problematic gaming and calls for action. The U.K. Chief Medical Officers' commentary on \enquote{Screen-based activities
and children and young people's mental health and psychosocial
wellbeing} has called on the technology industry to \enquote{develop
structures and remove addictive capabilities' from their services}.
Meanwhile, the World Health Organization has included gaming disorder in
the 11th Revision of the International Classification of Diseases
(ICD-11), defining it as a pattern of gaming behavior
(\enquote{digital-gaming} or \enquote{video-gaming}) characterized by \enquote{impaired
control over gaming, increasing priority given to gaming over other
activities to the extent that gaming takes precedence over other
interests and daily activities, and continuation or escalation of gaming
despite the occurrence of negative consequences}.\footnote{\enquote{ICD-11:
International Classification of Diseases 11th Revision -- 
The global standard for diagnostic health information}, WHO, 2019-05-25:
\url{https://icd.who.int/en}}

A number of authors note the importance of media literacy for children
and parents, together with parental supervision, to mitigate risks
stemming from online interactions (e.g.~\cite{griffiths2015adolescent}).
However, studies reveal that these approaches may reduce risks
-- without eliminating them~\cite{steeves2008closing}, vary
substantially in effectiveness depending on the type of parental
mediation and the age group of the children~\cite{lwin2008protecting}, and
suffer from increasing privacy and consent fatigue, which is promoted by
the opaqueness of online systems~\cite{hargittai2016can}. It is
argued that specifically this opaqueness, leading to \enquote{consent overload,
information overload, complexity of data processing, and lack of actual
choice,} also renders parental consent a highly questionable basis for
the lawful processing of children's data and an inadequate choice for
the protection of children's fundamental rights (cf.~\cite{simone2018importance}). In the following we discuss the ethics and harmful impact
of digital manipulation; we then revisit legal challenges, such as using
consent as a basis for data
processing.

\subsection{Neuromarketing and Ethics}
\label{sec:2-1-neuro}

\begin{quote}
\enquote{The mind, unconquered by violent passions, is a citadel, for a
man has no fortress more impregnable in which to find refuge and remain
safe forever}
-- Marco Aurelio, \emph{A sé stesso}
\end{quote}

The harms of digital manipulation impact on ethical aspects too.
Increasingly, companies are developing the ability to unobtrusively
observe and predict neurological states and functions, as well as to
manipulate emotions~\cite{zuboff_surveillance_2019}. \enquote{Stealthy
neuromarketing}~\cite{murphy_neuroethics_2008}, the manipulation of oblivious consumers with
the intent to addict and profit, limits their \emph{free will} using
them as means to an end, violating at once Rawlsian as well as Kantian
ethics~\cite{greene_for_2004, wilson_neuromarketing_2008}. As
behavioral science, neuro-science and deep learning techniques
increasingly confirm the substantial impact of biology on decision
making and action~\cite{fukuyama_our_2003, durante_ovulation_2011}, they are also
sabotaging the Renaissance concept of human autonomy; the use of
stealthy neuromarketing applied to children in gaming puts into question
the respect of fundamental ethical notions of
agency~\cite{wilson_neuromarketing_2008}, privacy~\cite{brownsword_regulating_2012}, freedom of
thought~\cite{mccarthy-jones_autonomous_2019} and mental integrity~\cite{ienca_towards_2017}.

Understanding the exact point where acceptable persuasion becomes
unacceptable manipulation is one of the crucial issues for the
regulation of digital manipulation~\cite{alegre_protecting_2021}. Luckily, the
literature on marketing ethics on the topic of persuasion versus
manipulation in advertising is lively and abundant~\cite{de_jans_advertising_2019}. Among many others, some authoritative views are represented by
Crisp~\cite{crisp_persuasive_1987}, who blames amorality on all forms of advertising
that override the autonomy of consumers; Arrington, who makes it a
question about what level of persuasion is the standard person assumed
to withstand~\cite{arrington_advertising_1982}; and Aylsworth, who allows for
manipulations, but only for ends that people accept and by means they
endorse~\cite{aylsworth_autonomy_2020}. As the theoretical debate develops, some tried
to create practical methodologies to distinguish ethical from unethical
persuasions~\cite{baker_tares_2001}, but new research is needed to
keep pace with the digital evolution~\cite{clarke_digital_2012}.

As the debate follows on, unethical combination of tactics manipulate
users toward the least privacy friendly options, jeopardizing the
genuineness of consent~\cite{council_deceived_2018}. The public and academic
skepticism~\cite{fischbach_why_2011} about uses of manipulating
techniques~\cite{wardlaw_can_2011} has called upon governments and market
regulators to act, as well as upon industry to adopt ethical guidelines,
etc~\cite{murphy_neuroethics_2008}. Manipulating children for
commercial purposes and enslaving them to addictive content is not the
type of digital future that society should wish for, but one that
discredits human dignity, creativity and diversity of thought. The
inviolable freedom of thought and opinion is described as \enquote{the
foundation of democratic society}, \enquote{the basis and origin of all other
rights}, without which freedom to think for ourselves \enquote{we lose our
freedom to be human}~\cite{alegre_regulating_2021}.

\section{Legal Challenges}
\label{sec:3-legal}

Considering the severity of effects (see
Section~\ref{sec:2-effects}), it is
surprising that there still are freemium games with manipulative and
addictive features available to children on easily accessible stores.

In such a complex context, the causes to the problem are various in
nature (technical, commercial, social, etc.) and intertwined (see
Sections~\ref{sec:1-1-paper} and~\ref{sec:2-1-neuro}).
It is unquestionable that the law, as one of the main means of
regulation~\cite{lessig_code_2009}, did have an impact on the current situation,
so what is important to understand is \emph{what} caused the legal
system to fail at children protection, and \emph{how} to fix it. As
there is no specific, unitary study on the matter, this paper is an
attempt to fill this gap.

Some causes of the failure might be rooted in the \emph{vexata quaestio}
of the efficacy of regulation based on risk together with notice and
consent mechanisms, especially in regards to parental
consent~\cite{gilbert_child_2011, livingstone_conceptualising_2018,
verdoodt_childrens_2020, livingstone_conceptualising_2018}. A substantial
portion of the legal scholarship is skeptical about the efficacy of
privacy notices (for a comprehensive review on the topic
cf.~\cite{van_den_berg_what_2012}, but also~\cite{barocas_notice_2009,
sloan_beyond_2014, acquisti_nudges_2017, macenaite_consent_2017,
hartzog_privacys_2018, schaub_design_2015, ben-shahar_more_2014})\footnote{Data protection became the law of
  everything which is \enquote{impossible to maintain}~\cite{purtova_law_2018} -- and
  enforce, and as an essential \enquote{right to a rule}~\cite{dalla_corte_right_2020}
  its enforcement depends on the interest to uphold the rule -- so long
  as the subject to it understands why the rule is there in the first
  place. Larose and Rifon~\cite{larose_promoting_2007} bring about the idea
  of data protection as protection against risks to data subjects'
  \emph{safety}. To them, the problem is one of \enquote{personal safety
  protection} as they critic the inadequacies of privacy policies
  models that do not motivate \enquote{consumers to take protective
actions}.
  Data subjects are like consumers who need means to \enquote{consider the
  potential consequences [\ldots] associated with personal information
  disclosures}, so that \enquote{they can make informed choices and enact
  appropriate behaviors that will shield them from online privacy
  threats}.}. Authoritative champion of this faction, Sartor claims
that data subjects' opportunity to consent to risks they cannot foresee
is not an \emph{asset} for them, but a \emph{liability}~\cite{sartor_regulating_2021}.\footnote{Sartor et \emph{al.}
  believe that \enquote{consent has been abused as a legal basis for targeted
  advertising}~\cite{sartor_regulating_2021}, p.~20: \enquote{The data
  subjects' power to consent or object to the processing of their
  personal data cannot be described as an asset -- a power to determine
  how they want their data to be processed -- but rather becomes a
  liability, something that makes them liable to surrender to any
  request made by businesses and platforms they interact with}.
  Similarly, the Article 29 working Party, in their \enquote{Guidelines on
  consent under Regulation 2016/679} state that: \enquote{If obtained in full
  compliance with the GDPR, consent is a tool that gives data subjects
  control over whether or not personal data concerning them will be
  processed. If not, the data subject's control becomes \emph{illusory}
  and consent will be an invalid basis for processing, [\ldots]}. European
Data Protection Board, 2020-05-04:
  \url{https://edpb.europa.eu/sites/default/files/files/file1/edpb_guidelines_202005_consent_en.pdf}}
Moreover, abundant research in the fields of behavioral decisions,
behavioral economics and experimental psychology applied to privacy by
Acquisti \emph{et al.} individuated privacy decision making hurdles that
affect -- if not altogether invalidate -- regulative efforts based on
the \enquote{privacy calculus}~(\emph{e.g.}, rules on consent as a legal basis
for data processing). They claim that incomplete and asymmetric
information, heuristics and bounded rationality, and cognitive and
behavioral biases~\cite{acquisti_nudges_2017} on the side of the data subject
mine the foundations of the \enquote{privacy calculus}~\cite{laufer_privacy_1977}, the theory of individual rational choice positing that data
subjects are agents with stable preferences in regards to privacy.

As for newer, less debated causes of the failure, the following are
worth developing.

One cause could be that existing regulatory frameworks, although based
on risks, are not receptive of the scientific evidence of risks and
consequences of manipulative and addictive content~\cite{sax_getting_2021}. It seems that there is a lack of subsumption of health, social
and psychological knowledge about gaming disorders into law, and the
problem might relate to multiple causes. On one hand, effects on
children will only be visible in many years; on the other, only the most
visible ones will be measurable -- think of incidence on rising gambling
practices, or gaming disorders that need medical attention --, whereas
\enquote{softer} effects, such as lack of compassion due to scarce critical
thinking, radicalization, devaluation of relationships, normalization of
gambling, and so on, will be harder to prove. Moreover, companies, such
as Facebook, who are best suited to study the effects of their services
on children, might cover their research showing that one of their most
famous services, that is Instagram, is \enquote{toxic} to teenage
girls.\footnote{Cf. \enquote{Facebook Knows Instagram Is Toxic for Teen
Girls, Company Documents Show}, The Wall Street Journal, 2021-08-14: 
  \url{https://www.wsj.com/articles/facebook-knows-instagram-is-toxic-for-teen-girls-company-documents-show-11631620739}}

Another cause could be that the applicable legal framework is too
complex and incoherent (cf. Table~\ref{tab:applicable-eu-law}).
It seems that the increasing addition of gaming features for attention
capturing~\cite{owen_is_2013, blades_advertising_2014,
van_reijmersdal_effects_2010, waiguny_relationship_2014} might have broadened the
regulatory framework of online gaming so that it became too complex to
be effective~\cite{sax_getting_2021, livingstone_protection_2018}. If games
might trigger regulation by consumer rights, data protection,
audiovisual media, marketing, digital content, products safety, national
gambling, health and criminal law, there is a chance that
over-regulation is jeopardizing rights
enforcement.\footnote{Regulation should have adapted to
  the rapid changes in the \emph{medium} without letting the
  responsibilities and protections of "this inter-disciplinary concern
  that often falls in between established mandates of relevant
  authorities" (cf. CoE Declaration) to be scattered. Moreover,
  "despite the growing importance of embedded advertising, many of the
  new ad formats remain neglected (e.g. native and mobile
advertising)~\cite{de_jans_advertising_2019}.} The latter is likely the case in the EU,
where a report from the Irish Council for Civil Liberties shows that
Europe has an \emph{enforcement paralysis}, and is \enquote{unable to police how
big tech firms use people's data.}\footnote{Cf. \enquote{Europe’s
enforcement
paralysis:
ICCL’s 2021 report on the 
enforcement capacity of
data protection authorities}:
  \url{https://www.iccl.ie/digital-data/2021-gdpr-report/}}

In response to problems of children protection, academia, international
organizations and civil society\footnote{E.g., UNICEF, 5rights Foundation,
Center for Humane Technology, EUkidsonline.net, and more.} have asked the
industry, regulatory authorities and policy makers to resort to actions.
The request to action vary from identifying and grading impacts of
persuasive design features, to creating a \enquote{fair game} charter with
\enquote{ethically child-centric standards}~\cite{murphy_neuroethics_2008, kidron_disrupted_2018,
livingstone_childrens_2019}, through
taking a public health approach,\footnote{Cf.
\enquote{Video games are pushing children into gambling warns
England’s top mental health nurse}, Health Tech Digital, 2020-02-19: \url{https://www.healthtechdigital.com/video-games-are-pushing-children-into-gambling-warns-englands-top-mental-health-nurse/}}
or creating new privacy preserving technologies and standards.\footnote{Cf.
the work of IEEE Standards Association presented in their \enquote{Webinar
Series:
Children's Data Governance
Applied Case Studies} at
\url{https://engagestandards.ieee.org/childrens-data-gov-webinar-register.html}}

In the EU, only the U.K. government has reacted to these requests by
approving the Age Appropriate Design Code, a set of 15 flexible
standards to which all organizations shall conform to as of September 2,
2021. Conforming to the code is necessary to comply with data protection
laws and should \enquote{ensure that an organization providing online services
likely to be accessed by children in the UK takes into account the best
interests of the child.}\footnote{\enquote{Age appropriate design: a code
of practice for online services}, The U.K.'s Information
  Commissioner's Office, available at
  \url{https://ico.org.uk/for-organisations/guide-to-data-protection/ico-codes-of-practice/age-appropriate-design-a-code-of-practice-for-online-services/executive-summary/}}
The standards are enforced on a proportionate and risk-based approach,
thus they do not prescribe or ban specific manipulative tactics;
however, concepts such as the \enquote{detrimental use of data} standard are
definitely worth exploring in upcoming policy making efforts.

Some of the recent proposals for future EU Regulations might also bear
some positive effects (see Table~\ref{tab:applicable-eu-law}:~\enquote{Potentially applicable EU laws
and provisions}), especially the finalized text of the proposal for an AI
Act which, in art. 5 sec. b)
specifically prohibits \enquote{the placing on the market, putting into service
or use of an AI system that exploits any of the vulnerabilities of a
specific group of persons due to their age, physical or mental
disability, in order to materially distort the behavior of a person
pertaining to that group in a manner that causes or is likely to cause
that person or another person physical or psychological harm.}
Unfortunately, the legislative \emph{iter} for this proposal is far from
finished; moreover, there is already authoritative criticism about
the AI Act by the European Data Protection Board and the European Data
Protection Supervisor jointly, as well as by Malgieri and Ienca, the
latter noting that the AI Act fails at protecting the \enquote{mind} of data
subjects by forgetting to \enquote{explicitly include in the high-risk list the
AI systems that rely on mental information such as emotion recognition
systems (in any form) and digital nudgers.}\footnote{Cf. \enquote{The EU
regulates AI but forgets to protect our mind}, European Law Blog,
2021-07-07:
  \url{https://europeanlawblog.eu/2021/07/07/the-eu-regulates-ai-but-forgets-to-protect-our-mind/}}

As a conclusive remark, and notwithstanding the inaction of most
European countries, it is not guaranteed that once the mentioned legal
causes are fixed, the bigger problem of children's digital manipulation
will be automatically solved too. In fact, studies on policies
versus gaming addiction by Király et al. show that
countries that are trying to address gaming problems through different
classes of policy attempts are failing at it. Measures limiting
availability of video games (e.g., shutdown policy, fatigue system, and
parental controls), reduce risk and harm (e.g., warning messages), or
help services for addicted players have proven to be
\emph{ineffective}~\cite{kiraly_policy_2018}. The reason, the authors say,
may be that such measures and policies \enquote{only addressed or influenced
specific aspects of the problem instead of using a more integrative
approach.} Therefore, the question: what could this \enquote{more integrative
approach} be? In the following sections we make two complementary
proposals to address this question, one technical and one legal.

\begin{table*}
\begin{center}
\caption{Potentially applicable EU laws and provisions}
\label{tab:applicable-eu-law}
\begin{tabular}{p{.35\textwidth} p{.55\textwidth}}
\hline
\textbf{Topic} & \textbf{Legislative Act}\tabularnewline
\hline
Information Society and Media & Audio Visual Media Service Directive
(amended, 2018)\tabularnewline
\hline
Judiciary and Fundamental Rights & Charter of Fundamental Rights of the
European Union (\enquote{Charter})\tabularnewline
& European Convention on Human Rights (CoE)\tabularnewline
& UN Convention on the Rights of the Child ("UNCRC")\tabularnewline
& Universal Declaration of Human Rights (UN)\tabularnewline
& International Covenant of Civil and Political Rights
(UN)\tabularnewline
\hline
Information Society and Media / Justice, Freedom and Security &
Directive combating the sexual abuse and sexual exploitation of children
and child pornography\tabularnewline
& Convention 108 (+) (CoE)\tabularnewline
& General Data Protection Regulation\tabularnewline
& ePrivacy directive\tabularnewline
& Unfair Commercial Practices Directive\tabularnewline
& Digital Content Directive\tabularnewline
& Consumer Rights Directive\tabularnewline
& Proposal for Regulation of: Artificial Intelligence, ePrivacy, Digital
Services, Digital Markets, Data
Governance\tabularnewline
\hline
\end{tabular}
\end{center}
\end{table*}

\section{Tech Support for Auditably Privacy-Preserving Gaming Platforms}

In the simplest terms, the ability to create personalized addictive
content from the side of the game provider is based on their ability to
collect and process fine-granular player data. In this section we focus
on a development approach that allows for games to be designed so as to
allow users to interact with the software, without the game provider
being able to tap into the interaction. Then, the provider would not be
able to tailor more targeted content.

Across the space of privacy-preserving technologies, researchers and
industry made a range of proposals, and built scalable infrastructures
for storage and processing of data with strong confidentiality
guarantees. For example, end-to-end encrypted cloud storage allows users
to store data on centralized infrastructure without the infrastructure
operator being able to access this data. Similarly, end-to-end encrypted
messaging services allow users to interact without the provider to
listen in on the exchanged information. On top of that, messaging
services with a focus on privacy, such as Signal~\cite{signal_signal_2021}, take
this a step further and hide the social graph of communicating user from
the infrastructure and service provider, in addition to also protecting
message confidentiality.

\subsection{Confidential Computing}

Technically this notion of privacy-preserving confidential messaging is
achieved by leveraging concepts of \emph{confidential computing}. In the
case of Signal, for example, TEEs~\cite{maene_hardware-based_2017} are used to
protect the service-side of the contact discovery mechanism from
interference by compromised infrastructure or from the service
provider~\cite{moxie0_technology_2017}: the service has published a contact discovery
protocol that is designed to not reveal a user's social graph to the
service by relying on irreversibly hashed user identifiers only. The
privacy-preserving properties of this protocol also rely on the correct
implementation of its service-side. To establish trust into this
implementation, Signal open-sourced the implementation and reproducibly
links the source code to the compiled service, and is executing the
resulting code in an \emph{enclave}\footnote{The term
  \enquote{enclave} was coined by Intel to denote isolated software execution
  in untrusted environments (e.g., on other people's computers), where
  hardware-based isolation mechanisms guarantee that potentially
  compromised system software cannot interfere with the integrity and
  confidentiality with code execution inside the enclave.} in a
server-side TEE. A client can then securely connect to the enclave and
obtain cryptographic proof that they are indeed interacting with the
intended remote software through a process called \emph{attestation}.
All this happens before the user transmits their hashed contacts to the
service, to establish a prior notion of trust. The verifiable guarantee
for the user, which results from the protocol briefly outlined above, is
that the Signal service persists no knowledge of a user's social graph.
Therefore the service cannot abuse or impart such knowledge for personal
gain, or if it is compromised by an attacker or
subpoenaed.

\subsection{Zero-Knowledge Gaming}

In this paper we outline and discuss a proposal to use technologies
similar to the \enquote{enclaved contact discovery} in the gaming sector. In
our vision, online games should be designed to minimize the
possibilities to collect or extract personal data, including non
personal data that might become personal through extensive processing
and aggregation. This can be achieved by compartmentalizing a game into
several components, which are then individually protected by means of
enclaved execution. Such a compartmentalization should depend on the
extent to which a component interacts with the user, processes personal
data, but may also reflect whether these components contain critical
intellectual property of the game producer. Privacy threat modeling
techniques such as LINDDUN~\cite{wuyts_linddun_2020} can help with
making privacy-conscious choices in this compartmentalization process.
The resulting components can then be individually audited and assessed,
and depending on their sensitivity, be enclaved as necessary. This would
lead to a notion of \emph{zero-knowledge
gaming},\footnote{Zero-knowledge gaming is a reference
  to \enquote{zero-knowledge proofs}, a group of cryptographic protocols that
  allow a party to prove knowledge of a secret to another party without
  revealing that secret.} where a gaming service may gain very little
knowledge about how users are using the service. The strategy for a user
to acquire trust in a game can then proceed, like in the case of the
Signal messenger, from a game component on the user's device which
attests the trustworthiness and integrity of remote components before
passing on data that could be used for extended user profiling. Ideally,
all such sensitive components should be open-source, be independently
assessed not to store or leak sensitive data, and be reproducibly built
from the assessed source code. Sensitive game components may, however,
pass pseudonymized, anonymized, or otherwise aggregated user data to
less sensitive game components. We believe that such a strategy does not
hinder the development of interesting game content, but makes
information flows explicit and reproducible for interested communities
or designated authorities, and verifiable by the users -- under the
assumption that they want to trust their devices and the locally
executing game components.

\subsection{A Concrete Example: Shufflepuck}

To give a brief example of such a compartmentalization strategy for a
mobile simulation of the game \emph{table
shuffleboard}.\footnote{Computer simulations of this
  game have a long history. More experienced readers might remember the
  famous Shufflepuck Cafè (cf.
  \url{https://en.wikipedia.org/wiki/Shufflepuck_Cafe}).}
Two players push simulated weights down a long table, aiming to hit a
scoring area at the opposite end of that table, which is defended by the
opposing player. For this game, we would consider two software
components:
\begin{paraenum}
  \item{a mobile app, which communicates with}
  \item{a server component.}
\end{paraenum}
The server would receive user inputs (shooting
angles, force, defensive actions) from the players' apps, it would then
compute the respective outcomes of the players actions, and communicate
these back to the apps, which visualize the game and process inputs from
the mobile devices' sensors. All software components -- the app as well
as the server -- could technically profile users and use profiling data
for purposes not intended by the users. This could be basic information
about players' contact details and who likes to play with whom (a social
graph!) but also delicate information such as someone's playing skills,
reaction times, and their behavior when they win or lose. In our model,
a core component of the game app would be assessed and reproducibly
built to only communicate user inputs (encrypted and
integrity-protected) to an attested server component. That server
component would be independently audited to not persist or leak the
information received. It might very well propagate a high-score table or
similar aggregated information. Also other game components, e.g. for
visualization, can receive aggregated, pseudonymized or anonymized
information from the core components, allowing for intellectual property
to be protected in these components.

The essential property of our proposal is that personal information
flows are being made explicit, auditable, and verifiable from the user's
device. In games where players interact or compete, personal data needs
to be used to distinguish individual players, to associate these players
with natural persons, and keep scores.\footnote{We
  assume basic features that require personal data to be essential in
  some games: One wants to play with or against one's friends,
  competitive players may want to be associated with their achievements,
  and a gaming economy may require player-specific financial
  transactions.} We argue that such data accumulation can be rendered
independent of in-game data, without losing reproducibility. As such,
our proposal dramatically reduces the possibilities for a mobile game to
record, store, and proliferate in-game data without this being flagged
during an audit or being reported as an integrity breach in a game
component. On the side of mobile apps, users already enjoy a certain
level of protection from malicious apps and spyware, as these are
(automatically) audited by mobile OS vendors.\footnote{App
  audits mobile OS vendors are not always effective and there are many
  potentially harmful apps available at app stores,
cf.~\cite{chatterjee_spyware_2018, suleman_combating_2021}.} The remote component of online games
is currently not part of such audits, which our proposal strives to
address.

\subsection{Limitations \& Challenges}

Of course, our design has a few drawbacks. For example, game developers
may be scared by the additional complexity introduced by
compartmentalization and TEE features. Ongoing work~\cite{scopelliti_poster_2021} aims to address this by providing cross-platform abstractions to
build distributed TEE applications. We are critically aware of the
dangers of working with (often proprietary) hardware-based TEE
platforms. Proposals to provide similar security and privacy guarantees
in the context of discovering social graphs but without trusted hardware
do exist (e.g.,~\cite{demmler_pir-psi_2018, kales_mobile_2019}, yet, these have
not been implemented in mainstream messaging services. A range of
attacks against hardware security frameworks that may compromise the
security of a contact discovery approach such as Signal's have been
published (e.g.,~\cite{van_schaik_sgaxe_2020}) and we deem research on securing
processors and cryptographic algorithms as orthogonal to our proposal.
Ongoing work in the space of homomorphic encryption, zero-knowledge
protocols, and secure multi-party computations may be capable of
providing the confidentiality and authenticity guarantees of TEEs
without relying on specific hardware support.

Remaining issues may, however, be with performance and scalability for
certain highly interactive and computationally intensive games. In the
case of games for smaller children, where additional transparency and
enforcement of data protection may be most needed, we do not see such
constraints. We further allow for a notion of end-to-end encryption in
multi-player online games, which makes it impossible for the gaming
service or third parties to trace fraud (e.g., in-game cheating or
fraudulent player item trading between players) or abuse (e.g., verbal
abuse or grooming through an in-game chat). Thus, games must be designed
not to have features that can lead to fraud or abuse. Ultimately, we
believe that user's trust is a key element of a privacy-preserving
gaming platform but, regrettably, it is no easy task to gain trust in
such a platform. In our case, certain properties of a software product
are verified at run-time by means of cryptographic methods -- this
happens, however, between machines (e.g., complex computations on the
user's smartphone and cloud infrastructure) and is not immediately
reproducible by the user. Therefore, the trust-establishing element of
such a privacy-preserving gaming system can only be the independent and
repeated audit of a game (or a gaming platform, if such a generalization
is feasible) by a dedicated community or authority, who follows a strong
regulatory framework that requires data minimization and that clearly
defines \enquote{labels} to tag games that deviate from the data minimization
requirement in whatever way. Most importantly, however, introducing such
a regulatory framework and the technological backing to enforce the
regulations, will require a shift in business models for many
free-to-download games that are based on targeted advertising.

We have been explaining the idea behind our approach based on a simple
simulation game, which may not be considered representative for games
typically played by today's teenagers. Importantly, and depending on
implementation specifics, our example may very well feature the full
scale of game analytics and player metrics of a \emph{jump `n' run} game
or a complex multiplayer role-playing game because player inputs
interactions may be of similar complexity, and the goals of the gaming
company, that is maximizing revenue through attention (Alha 2020), are
not affected by the complexity of the game
(cf.~\cite{seif_el-nasr_introduction_2013}).\footnote{In fact, several of the
  highest-grossing games a puzzle games with remarkably simple game
  mechanics:
  \url{https://en.wikipedia.org/wiki/List_of_highest-grossing_mobile_games}} Most importantly, our idea does not actually restrict player
interactions and game dynamics. Instead, we propose data flows between
game components to be made explicit and distinguishable, ensuring that
an auditor can assess these data flows and that a game company cannot
change critical parts of a game without this being detectable by users.
As such, our proposal may ask for new privacy-centered game
architectures that have not been explored to date. We aim to follow up
on these questions in experimental work in the future, where we explore
different software architectures, aiming to refactor a representative
open-source game to implement our idea.

After having discussed the technological side of the integrative
approach, we now move to the legal
aspect.

\section{Children's Right to Freedom of Thought}

The call for a more integrative approach is getting traction in both
academia~\cite{alegre_regulating_2021, kidron_disrupted_2018,
acquisti_nudges_2017} and
international organizations.\footnote{Cf. e.g.,
  CoE~\cite{council_of_europe_technological_2017} and the Declaration, and the United
  Nations' Secretary-General's High-level Panel on Digital Cooperation
  and the subsequent Secretary-General's Roadmap for Digital Cooperation
  and reports by David Kaye, former UN Special Rapporteur on the
  promotion and protection of the right to freedom of opinion and
  expression} Already in 2017, with the Recommendation 2102
(2017)~\cite{council_of_europe_technological_2017}, the CoE asked for guidelines on \enquote{the
recognition of new rights in terms of respect for private and family
life, the ability to refuse to be subjected to profiling, [\ldots] and
\emph{to be manipulated or influenced}.} In the 2019 Declaration,
the CoE pointed out that European policy has focused and relied
intensively on people's protection through personal data regulation,
while instead it should have moved \enquote{beyond data protection}. As a
result, the search for \emph{actual} impacts of digital manipulation
through digital technologies and AI\footnote{There is a
  \enquote{need for additional protective frameworks related to data that go
  beyond current notions of personal data protection and privacy and
  address the significant impacts of the targeted use of data on
  societies and on the exercise of human rights more broadly}~(Europe
  2019)} is currently a key issue for the CoE Ad hoc Committee on
Artificial Intelligence (CAHAI).

Meanwhile, from the perspective of law in practice, early forms of
recognition of addictive technologies are starting to appear. For
example, the case \emph{Beau Zanca, et al. v. Epic Games, Inc.} is the
first American class action against the gaming company Epic Games. In
this case, the maker of the blockbuster freemium Fortnite reached a
settlement to avoid the chance of being found guilty of wrongdoing over
the use of loot boxes in children's games. As a second example, another
class action \emph{F.N. et J.Z. v. Epic Games Inc. et al.} has been
filed by Calex legal in Quebec, Canada, on behalf of two parents of
minors lamenting \enquote{the highly addictive nature of the game.} The
development of this case will be particularly relevant for European
consumers in view of the entering into force of the Directive EU
2020/1828 on \enquote{representative actions for the protection of the
collective interests of consumers.}

\subsection{Beyond Privacy}

The current legal literature on digital harms argues whether existing
right to privacy and its interpretations are enough to face the new
challenges posed by AI, big data analytics and neuromarketing. To some,
secondary law implementations in the EU (mostly GDPR) and latest
interpretations of risks by the European Data Protection Board can
withstand such challenge today~\cite{ienca_mental_2021}.

However, it is increasingly getting traction the idea that EU regulation
failed to recognize that people's protection from digital harms cannot
always be granted by only respecting rules on data processing; this is
because, as is the case of children's protection from manipulation,
sometimes the context changes so much that the ultimate human right at
risk is no longer digital privacy~\cite{alegre_regulating_2021}, but something beyond
it.

Some believe that to move beyond privacy means the need for new
rights~\cite{skriabin_neurotechnologies_2021, sieber_souled_2019,
ienca_towards_2017}.
Above all, Ienca and Adorno claim that the implications raised by
neuroscience and neurotechnology \enquote{urge a prompt and adaptive response
from human rights law with new neuro-specific
rights}~\cite{ienca_towards_2017}.\footnote{Although they individuate four
  rights to cognitive liberty, mental privacy, mental integrity and
  psychological continuity, only that to cognitive liberty seems
  relevant to develop a theory of mental self-determination and freedom
  from manipulation (\emph{see infra}).}

As a different approach, that is moving beyond privacy without creating
new rights, Alegre calls upon a focus shift from privacy to \enquote{freedom of
thought}~\cite{alegre_protecting_2021}, which seems to be also the approach that the
CoE and the UN are interested in developing. In fact, the Committee on
the Rights of the Child, in General Comment 25 to the UNCRC is the
\enquote{first clear articulation of the right to freedom of thought in the
digital age from a UN body}~\cite{alegre_protecting_2021}, as it
\enquote{encourages States parties to introduce or update data protection
regulation and design standards that identify, define and prohibit
practices that manipulate or interfere with \emph{children's right to
freedom of thought} and belief in the digital environment, for example
by emotional analytics or inference.}\footnote{The
  quote so continues: \enquote{Automated systems may be used to make inferences
  about a child's inner state. They should ensure that automated systems
  or information filtering systems are not used to affect or influence
  children's behaviour or emotions or to limit their opportunities or
  development.}}

We believe that the approach of Alegre, the CoE and the UN to shift from
privacy to freedom of thought is the right one to adopt. We discuss the
reasons for the \enquote{upgrade} of the right to privacy to that of freedom of
thought in the following sections.

\subsection{Regulating From the Perspective of Freedom of Thought: Ancient
Rules in Future-Proof Regulation}

\begin{quote}
\enquote{Regulating from the perspective of the right to freedom of
thought is new and complex, but it is crucial to our future as
autonomous humans living in democratic societies founded on human
rights} -- Susie Alegre~\cite{alegre_protecting_2021}
\end{quote}

Old and dusted, the right to freedom of thought has been shadowed by the
right to privacy for the past 15 years and now needs new interpretations
in the light of the digital evolution. In its digital clothes, freedom
of thought is best described in the literature by the concept of
\enquote{cognitive liberty}~\cite{ienca_towards_2017, bublitz_my_2013,
sententia_neuroethical_2004}. It should be understood as the
\enquote{conceptual update}~\cite{sententia_neuroethical_2004} of the right to privacy in its self-determination connotation,
mixed with interpretations of freedom of thought~\cite{alegre_protecting_2021} already
in the jurisprudence of the European Court of Human Rights (\emph{see}
\emph{Nolan and K. v Russia}\footnote{\enquote{Case of Nolan and K. v.
Russia}, European Court of Human Rights, 2512/04: \url{https://hudoc.echr.coe.int/eng?i=001-91302}}), and as expressed in international human
rights legislation.\footnote{Article 19 of the
  International Covenant on Civil and Political Rights, The right to
  \enquote{hold opinions without interference} art. 18 of Universal
  Declaration of Human Right and art. 10 of the EU Charter~\cite{alegre_protecting_2021} \enquote{Cognitive liberty is the neuro-cognitive substrate of all
  other human and civil liberties,} clarifies the UN Human Rights
  Committee, and it is of utmost importance that it is constantly
  protected even in a precautionary fashion., On this point, the UK
  Chief Medical Officers recommends that, while new research is being
  carried out, technology companies \enquote{recognise a precautionary approach
  in developing structures and remove addictive capabilities.}}

Applying the right to freedom of thought to children's rights would not
be a new concept. Multiple laws in most western legal systems have long
since provided for that, dating back to the 2nd century BC, when the
\emph{Lex laetoria de circumscriptione adolescentibus} gave action
against who fraudulently induced a minor to enter into a transaction and
condemning them with \emph{infamia}, the loss of
reputation~\cite{candy_lex_2018}. Emanation of moral value, the general prohibition of the use of
manipulative tactics to take advantage of children's
vulnerability~\cite{okeeffe_impact_2011} has been early
embedded into roman contract and criminal law, and nowadays into EU
human rights, commercial, advertising, market, audiovisual, and digital
services laws. We are of the opinion that, today, it must be recognised
and defined specifically in human rights law, to make sure it does not
get lost in the unreliable protections of data privacy (see
Section~\ref{sec:3-legal}) or in the
complex application of the best interest of the child. This rule's
\emph{rationale} is the protection of the freedom of a child to develop
their self without interference from third parties that do not have
their best interest in mind, such as forms of commercial
exploitation~\cite{third_young_2017}. It is the children's right to be children,
to experience, to make mistakes, and to do so without malevolent
interference from adults.

\subsection{Challenges Ahead}

Notwithstanding the maturity of this ancient principle of law,
regulating today's technologies from the perspective of freedom of
thought implies radical substantial and procedural changes: whereas
privacy and data protection are non-absolute rights that have been
limited by balancing with other fundamental rights -- such as the right of others
to conduct business --, freedom of thought is an \emph{absolute} right
and does not accept any interference.\footnote{\enquote{Little
  has been done to develop fully operational legislative and regulatory
  frameworks to ensure that enjoyment of the right to freedom of thought
  is real and effective in a modern context [\ldots] there are also
  positive obligations to protect all those in their jurisdiction from
  interference with the right. So, laws must be in place, not only to
  prevent state actions that could interfere with our rights to freedom
  in the `forum internum,' but also prohibiting others from doing
  so.}~\cite{alegre_regulating_2021}}

To this day, the regulators have been addressing personalized addictive
content under the assumption that a well designed system of procedural
rules on personal data collection and processing could be the adequate
shield with which to protect all of data subjects' fundamental rights.
We instead claim otherwise: some of the cornerstones of EU data
protection, most notably its risk-based approach and its tolerance to
lawful interferences -- as in, \enquote{procedurally sound} interferences --
have allowed organizations to keep poking holes and finding cracks in
such shield, exploiting state-of-the-art technical and organizational
measures for the market-necessary aim of demonstrating compliance. We
instead believe that what's important is not so much whether procedural
data protection rules on processing are respected, but whether the
fundamental rights of children, which those rules aim to protect, are
respected in their essence. In practice, this means that when a
processing activity interferes with the right to children's freedom of
thought, notwithstanding how well it respects data protection rules, it
is a violation of such right.

A shift from data privacy to protection of freedom of thought means
changing the status of the fundamental right being protected, from
relative to absolute, which in turn will have systemic legal
repercussions, and calls for a rethinking of the entire regulatory
ecosystem: from its ethics, to human rights interpretations, to
adaptations of secondary law (see EU Legal Acquis, Table~\ref{tab:applicable-eu-law}),
and so on. All this considered, it seems that the biggest challenge
ahead, at least from a legal perspective, is to understand how
regulating digital manipulations through the right to freedom of thought
is going to affect the EU legal system, and what will be its
repercussions.

\section{Conclusions \& Future Directions}

The predatory tactics and tools of the attention economy have shown
their worst side in the context of gaming. Freemium business models are
harming children's health, social, and personal well-being, in a
fast-changing technological context where regulators have failed at
being receptive to the new risks, and to adequately respond with
effective regulatory means. We believe that if regulators keep
addressing the issue of children digital manipulation through new
regulation along the lines of established concepts involving data
protection and privacy, it will only incentivize novel analytics and a
diversification of addictive strategies rather than prioritize the best
interest, and uphold the fundamental rights of children.

We discuss reasons whether and why decision makers' reliance on European
privacy laws for the protection of children from digital manipulation
has proven unsuccessful. The failure, we claim, is not only due to
reasons of inefficient law enforcement systems (see Section~\ref{sec:3-legal} on the \enquote{enforcement
paralysis}), but also to a fundamental error in the detection of the
asset to protect. This asset, in fact, we do not identify as the
relative human right to privacy, rather the absolute right to freedom of
thought.

Pending new regulations, we propose one technical and one regulatory
approach to address this issue: a zero-knowledge model for a gaming
information system, to \enquote{dab the bleeding} with a technical fix, and a
regulatory \enquote{upgrade}, which is a shift from privacy to freedom of
thought. The interplay of the two should provide for an \enquote{integrative
approach} with immediate as well as long term effects, hopefully fit to
shape a more humane future that forbids children's manipulation and
exploitation for commercial purposes.

The idea of such an \enquote{integrative approach} needs to be picked up and
considered across regulatory bodies, authorities that oversee the
development of content for minors, technology companies, and researchers
alike. Specifically, we see a need for action by competent national
regulatory authorities for streamlining the analysis of manipulative
neuromarketing techniques (in view of the upcoming AI Act and its
classification of prohibited AI due to manipulative practices) and
homogenizing the assessment of online games. As for policy makers, we
believe they should invest resources into understanding the risks of
addictive technologies and finding regulatory solutions. Game
developers, on their side, should become aware of risks of addictive
technologies and make conscious choices about ethical software design,
as well as foster digital alphabetization and raise users' risk
awareness.

% \paragraph{Acknowledgments.}
% %

\begin{acks}
This research is funded by the Research Fund KU Leuven, and by the Flemish
Research Programme Cybersecurity. The research was also funded by the EU
project LeADS, GA 956562.

\end{acks}

%\newpage
\bibliographystyle{alphaurl}
\bibliography{main.bib}

%\clearpage
%\footnotesize
%\verbatiminput{20220430-reviews.txt}

\end{document}